\documentclass{desyproc}

\begin{document}
\title{A fresh look on the limit\\on ultralight axion-like particles from SN1987A}

\author{{\slshape Alexandre Payez}\\[1ex]
Deutsches Elektronen-Synchrotron (DESY), Hamburg, Germany}

\contribID{payez\_alexandre}

\confID{300768}  
\desyproc{DESY-PROC-2014-03}
\acronym{Patras 2014} 
\doi  

\maketitle

\begin{abstract}
We revisit the limit on very light axion-like particles (ALPs) from the absence of gamma rays coincidental with the neutrino burst from SN1987A.
We use updated supernova simulations, modern models for the magnetic field inside the Galaxy, and a Primakoff cross section which takes into account proton-degeneracy and mass-reduction effects.

We finally give an updated exclusion plot for the electromagnetic coupling of sub-eV ALPs, comparing our new bound with other limits as well as with future ALP searches.
\end{abstract}

\section{Reminder}
	
	Axion-like particles (ALPs) are generic predictions of theories beyond the Standard Model of particle physics, where they essentially arise as pseudo-Nambu--Goldstone bosons of new spontaneously broken global symmetries.
	The case for such light	particles is actually arguably getting even more interesting since the LHC discovered the scalar Higgs boson\hspace{1pt}---\hspace{1pt}and gave an experimental evidence of how important spontaneous symmetry breaking can be in particle physics\hspace{1pt}---\hspace{1pt} but has however so far found no new heavy particle beyond the Standard Model, and in particular no signs of SUSY where it was presumably expected.

	Together with the long-sought QCD axion, there is also some interest for ultralight ALPs, somewhat driven by phenomenology: indeed, a number of observations in astrophysics have been claimed by various authors to hint at the existence of such nearly massless (pseudo)scalars, thereby defining another window of interest in the ALP parameter space. The reason why they would be so interesting is the electromagnetic coupling that they might have, that would affect in a number of ways the signals expected from astrophysical sources~\cite{SikivieRaffeltHarariSikivie}.
	There of course exist strong constraints on the electromagnetic coupling of such light particles, and the aim of this work~\cite{Payez:2014} is actually to revisit and update what has remained for almost 20 years the most stringent bound over a wide range of masses in the astrophysical window~\cite{Grifols:1996id,Brockway:1996yr}.

	When a very massive star undergoes a core-collapse, lots of neutrinos are quickly radiated by the proto-neutron star, leading to a short and intense neutrino burst that will arrive at Earth hours before the optical flash.
	Such supernova (SN) explosions are in fact also an ideal place to search for extremely light ($m_a \lesssim 10^{-9}$~eV) ALPs $a$ with a generic two-photon interaction, of effective coupling $g_{a\gamma}$:
	\begin{equation}
		\mathcal{L}_{a\gamma\gamma}  = \frac{1}{4} g_{a\gamma} F_{\mu\nu} \tilde{F}^{\mu\nu} a.
	\end{equation}
	Produced with typical energies related to the core temperature via the Primakoff effect on $p^+$,
	\begin{equation}
		p^+ + \gamma \rightarrow p^+ + a,
	\end{equation}
	these light spinless particles would then quickly escape the exploding star, and later convert in the Galactic magnetic field into $\gamma$-ray photons, coincidental with the neutrino burst~\cite{Grifols:1996id,Brockway:1996yr}.

	Such a signal has been searched for in February 1987, when it was realised that the star Sanduleak -69$^\circ$ 202 had blown up as a core-collapse supernova in the Large Magellanic Cloud in an event known as SN1987A: it was the first supernova in more than 300 years that could be observed with the naked eye, located only $50$~kpc away from us. As predicted, a burst of neutrinos was detected; however, only an upper limit could be obtained on the $\gamma$ signal during this 10-s burst, leading to the well-known constraints on the coupling of light axion-like particles which we want to revisit: $g_{a\gamma} \lesssim 3 \times 10^{-12}$~GeV$^{-1}$~\cite{Grifols:1996id} or $g_{a\gamma} \lesssim 10^{-11}$~GeV$^{-1}$~\cite{Brockway:1996yr}, for $m_a \lesssim 10^{-9}$~eV.

\section{Main updates}

\subsubsection*{Supernova simulations \& time resolution}

The two original independent analyses have adopted different approaches: Ref.~\cite{Grifols:1996id} relied on analytical results only and considered typical estimates of the conditions inside a SN core, while Ref.~\cite{Brockway:1996yr} considered both analytical and numerical results, using simulations of the conditions inside the star at 1~s, 5~s and 10~s after the core bounce and eventually making a linear regression to estimate the integrated photon flux at Earth over this time interval.

In our work, we use recently updated spherically symmetric supernova models~\cite{Fischer:2009af} for two progenitor masses: $18~M_{\odot}$ (resp. $10.8~M_{\odot}$), with simulations up to $21$~s (resp. 10~s) after bounce. Our main result is obtained with the 18~$M_{\odot}$ progenitor, while the 10.8~$M_{\odot}$ one allows us to investigate the dependence of our limit on the progenitor mass.
In both cases, the simulation data consist of a collection of $\sim600$ snapshots at different times giving the profiles of various physical quantities inside the proto-neutron star as a function of the radius.
This also allows us to get a more precise timing evolution of the ALP production inside the supernova.

\subsubsection*{Magnetic field \& conversion probability}

We make a very significant improvement for the description of the magnetic field.
Indeed, the original studies simply considered a homogeneous Galactic magnetic field of transverse field strength $B_T \sim 1~\mu$G over $L = 1$~kpc, while we now use the recent model of Jansson and Farrar of the Galactic magnetic field~\cite{Jansson:2012pc}, and have also performed all our calculations using the one of Pshirkov \textit{et al.}~\cite{Pshirkov:2011um} for comparison.
A major difference between these elaborate models and the original toy model is that they take into account the presence of a halo component over several kiloparsecs; the various bounds that we obtain are therefore actually more stringent. 

Both the original papers~\cite{Grifols:1996id, Brockway:1996yr} also
used an approximate expression of the conversion probability, with $\Delta\mu^2$ being the difference of the mass eigenvalues squared, and $\theta$, the mixing angle:
\begin{equation}
	P_{a\gamma} = \sin^2(2\theta) \sin^2(\frac{\Delta\mu^2 L}{4E})	\sim	\frac{1}{4}\ g_{a\gamma}^2 B_T^2 L^2.
\end{equation}
For the case at hand, this approximation would be valid in the massless limit and was estimated to hold for $m_a \leq 10^{-9}$~eV. In our calculations, we instead now consider the full conversion probability and are therefore able to directly derive the mass dependence of the limit.

\subsubsection*{Degeneracy \& high density}

	We further include effects related to the conditions in the SN core that were mentioned but not included to calculate the ALP production in Refs.~\cite{Grifols:1996id, Brockway:1996yr}.
	The first one is to take into account the fact that the protons are partially degenerate. This affects the number of available targets and also the screening inside the plasma (somewhere between the Debye and the Thomas--Fermi regimes). The second is the fact that, due to the extremely high density during the first seconds ($\rho\sim10^{14}$~g~cm$^{-3}$), the $p^+$ effective mass can actually go down to about 50\% of its value in vacuum. This makes the protons easier to be degenerate but also means that there are more targets available for a given mass density. We further take this mass reduction into consideration, and use the updated tables of equation of state for nuclear matter in a supernova (2010, 2011) based on Ref.~\cite{Shen:1998}, which were actually also used in the SN models themselves~\cite{Fischer:2009af}.

\section{Bottom line}

\begin{figure}[h!]
\centerline{\includegraphics[width=\textwidth]{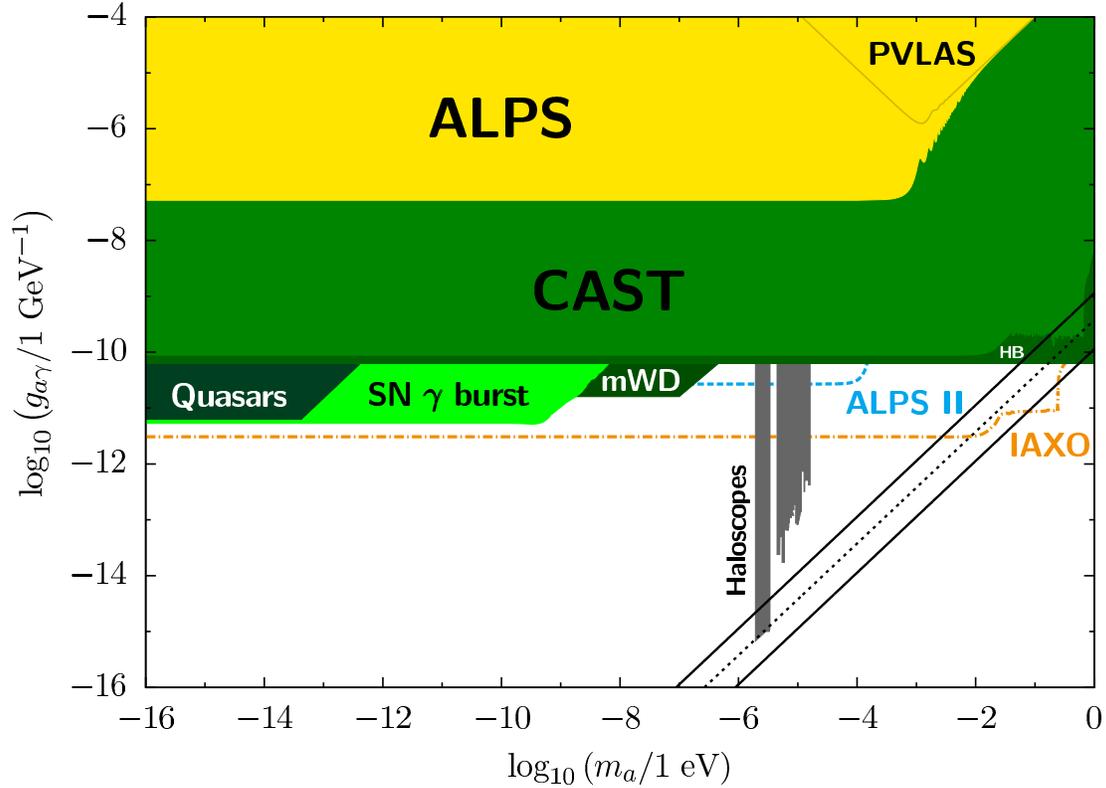}}
\caption{Update of the exclusion plot of Ref.~\cite{Cadamuro:2012rm} for sub-eV ALPs. We compare our updated bound with some other limits that apply in this region of the parameter space. We also indicate the projected sensitivities of the forthcoming ALPS II experiment~\cite{Bahre:2013ywa} and of the proposed IAXO experiment~\cite{Vogel:2013bta}.}\label{fig:exclusionplot}
\end{figure}

Putting together the improvements discussed above, the low-mass bound that we obtain is~\cite{Payez:2014}
\begin{equation}
	g_{a\gamma} \lesssim 5.3 \times 10^{-12}~{\rm GeV}^{-1},\quad {\rm for } \quad m_a \lesssim 4.4 \times 10^{-10}~{\rm eV}.
\end{equation}
The results are very stable over a variety of changes mostly because the limit on $g_{a\gamma}$ essentially goes as the fourth root of the fluence. 
We give an exclusion plot for low-mass ALPs in Fig.~\ref{fig:exclusionplot}, updating the one from Ref.~\cite{Cadamuro:2012rm} to include our main result, obtained using the model of Jansson and Farrar for the Galactic magnetic field.
We also compare the new limit with the constraint from quasar polarisation measurements~\cite{Payez:20122013} that corresponds to the magnetic field and electron density assumed in Ref.~\cite{Vallee:2002ms} in the plane of the local supercluster.

\section*{Acknowledgments}

We would like to thank all our collaborators on this project: Carmelo Evoli, Tobias Fischer, Maurizio Giannotti, Alessandro Mirizzi, and Andreas Ringwald.
It is also a pleasure to thank Davide Cadamuro for sharing the data he used to produce his own exclusion plot.
This work has been supported by the German Science Foundation (DFG) within the Collaborative Research Center SFB 676 ``Particles, Strings and the Early Universe.''


\begin{footnotesize}

\end{footnotesize}


\end{document}